\documentstyle[epsf,11pt]{article}
\newlength{\bredde}
\def\slash#1{\settowidth{\bredde}{$#1$}\ifmmode\,\raisebox{.15ex}{/}
\hspace*{-\bredde} #1\else$\,\raisebox{.15ex}{/}\hspace*{-\bredde} #1$\fi}
\newcommand{\beq}{\begin{equation}}
\newcommand{\eeq}{\end{equation}}
\newcommand{\noi}{\vspace{12pt}\noindent}
\newcommand{\lG}{\raise.3ex\hbox{$\stackrel{\leftarrow}{G}$}}
\newcommand{\lU}{\raise.3ex\hbox{$\stackrel{\leftarrow}{U}$}}
\newcommand{\lP}{\raise.3ex\hbox{$\stackrel{\leftarrow}{{\cal P}}$}}
\newcommand{\leta}{\raise.3ex\hbox{$\stackrel{\leftarrow}{\eta}$}}
\newcommand{\lOmega}{\raise.3ex\hbox{$\stackrel{\leftarrow}{\Omega}$}}
\newcommand{\ldr}{\raise.3ex\hbox{$\stackrel{\leftarrow}{\delta^r}$}}
\def\beqn{\begin{eqnarray}}
\def\eeqn{\end{eqnarray}}

\def\gtwid{\raise.3ex\hbox{$>$\kern-.75em\lower1ex\hbox{$\sim$}}}
\def\ltwid{\raise.3ex\hbox{$<$\kern-.75em\lower1ex\hbox{$\sim$}}}

\def\slash{\!\!\!\!/\,}
\def\Dslash{{D}\slash}

\begin{document}
\topmargin -1.4cm
\oddsidemargin -0.8cm
\evensidemargin -0.8cm
\title{\Large{Microscopic Spectral Density of the Dirac Operator
in Quenched QCD}}

\vspace{3cm}

\author{\\~
{\sc P.H. Damgaard$^a$, U.M. Heller$^b$ and  A. Krasnitz$^c$}
\\~\\~$^{a)}$ The Niels Bohr Institute\\Blegdamsvej 17\\DK-2100
Copenhagen\\Denmark\\~\\~$^{b)}$ SCRI\\Florida State University \\
Tallahassee, FL 32306-4130\\USA\\~\\~$^{c)}$Unidade de Ci\^encias
Exactas e Humanas\\Universidade do
Algarve \\ Campus de Gambelas, P-8000 Faro \\ Portugal
}
 
\maketitle
\vfill
\begin{abstract} 
Measurements of the lowest-lying eigenvalues of the staggered fermion Dirac
operator are made on ensembles of equilibrium gauge field configurations 
in quenched SU(3) lattice gauge theory. The results are compared with exact
analytical predictions in the microscopic finite-volume scaling regime.
\end{abstract}
\vspace{0.3cm}

\vfill
\begin{flushleft}
NBI-HE-98-33 \\
FSU-SCRI-98-120 \\
hep-lat/9810060
\end{flushleft}
\thispagestyle{empty}
\newpage

\noindent
The study of the eigenvalue spectrum of the Dirac operator in QCD and
related gauge theories has a long
history, most notably due to the relation between zero modes,
near-zero modes, topology, and chiral symmetry breaking. 
More recently the subject
has undergone a strong revival. This was originally sparked by the 
observation made by Leutwyler and Smilga \cite{LS} that in sectors of fixed 
topological charge $\nu$ one can derive exact analytical results for the 
distribution of small eigenvalues of the Dirac operator. 
The idea is that there 
exists a finite-volume scaling regime for the euclidean gauge theory where 
the partition function can be computed exactly. One crucial input is the
assumption that chiral symmetry is spontaneously broken according to
the conventional scenario. The finite-size scaling regime is 
unphysical in the sense that it restricts the associated pseudo-Goldstone 
bosons to be very light, so light that only their zero-momentum 
modes contribute
significantly to the euclidean path integral. The conditions for this
simplification are two-fold. First, the euclidean finite-volume partition 
function of QCD must be accurately described by the effective hadronic
excitations, rather than the underlying quark and gluon degrees of freedom.
This imposes the large-volume condition $1/\Lambda_{\mbox{\rm 
{\footnotesize QCD}}} 
\ll L$, where $L$ is the linear extent of the four-volume. Second, this
linear extent $L$ should always be insignificant compared with the Compton 
wavelength of the pseudo-Goldstone bosons: $L\ll 1/m_{\pi}$. While this 
latter condition
is inappropriate for actual, physical, no-so-light pions, it is of course
just the right limit for studying spontaneous chiral symmetry breaking.
More importantly in this context: it is ideally suited for finite-volume 
studies of lattice gauge theory.

\noi
A major breakthrough in the understanding of the Dirac operator spectrum
in the above finite-volume region has been the suggestion by 
Verbaarschot and co-workers that random matrix theory can give exact analytical
predictions for the spectrum in a suitably defined microscopic regime
\cite{SV,V}. Gauge theories based on various gauge groups and with 
fermions in either fundamental or adjoint representations fall into
essentially just three different universality classes, which beautifully
fit into corresponding classifications of random matrix theory ensembles
\cite{V1}. In this
paper we shall restrict ourselves to SU(3) gauge theory with quarks in
the fundamental representation of the gauge group, where universality of all
microscopic correlation functions, including the microscopic spectral
density itself, has been proven \cite{ADMN}. Moreover, it has 
recently been shown that all these exact results can also be derived 
directly from the finite-volume partition function of QCD \cite{D,OTV}, 
without recourse to random matrix theory. Also the universal 
smallest-eigenvalue distribution can be computed in this way \cite{NDW}. 

\noi
So far the analytical predictions for the microscopic Dirac operator
spectrum have been tested in 4-dimensional SU(2) lattice gauge theory with
staggered fermions \cite{BB}, and in 3-dimensional SU(3) lattice gauge
theory with staggered fermions \cite{DHKM}. There have also
been preliminary studies of the microscopic spectrum in 2-dimensional
U(1) lattice gauge theory with a fixed-point Dirac operator \cite{L}.  
However, none of these cases test the most important universality class of 
all, namely that of (in the language of random matrix theory) the chiral 
Unitary Ensemble. This is the universality class of 
4-dimensional SU(N$_c$) gauge theories with $N_c \geq 3$, and quarks
in the fundamental representation.

\noi
In this paper we shall present the first results for a lattice gauge
theory determination of the microscopic Dirac operator spectrum of
SU(3) gauge theory with a staggered fermion Dirac operator. The biggest
advantage of using a staggered Dirac operator is that the microscopic 
spectrum at any value of the gauge coupling should coincide with the
continuum predictions, since the universality class of this 
lattice-regularized Dirac operator in this case 
coincides with the universality class
of the continuum Dirac operator (for fermions
in the same color representation).
The main disadvantage is related to the fact that all
comparisons with analytical results should be done in sectors of fixed
topological charge $\nu$. It is well-known that although a remnant of chiral
symmetry is exactly preserved in the lattice theory with staggered
fermions, the interplay between gauge-field topology and fermions is a 
subtle issue in this formulation. In particular,
the existence of $\nu$ fermion zero modes in a gauge field
configuration with topological charge $\nu$ is jeopardized. Earlier
studies of the microscopic Dirac operator spectrum in SU(2) lattice gauge 
theory with staggered fermions \cite{BB} have suggested that to a large 
degree the effects of 
non-trivial gauge field topology are insignificant at feasible values
of the gauge field coupling, and hence can be ignored at this stage. While we
feel that this issue still deserves further clarification, we shall
here take the same stand as in ref. \cite{BB}, and simply sum over all
gauge field configurations, irrespective of topology. Comparison should,
in the same spirit, be made only with analytical $\nu=0$ predictions.  

\noi
Before considering the numerical algorithm that computes the Dirac
operator eigenvalues in a given gauge field configuration, it is instructive 
to estimate the number of small eigenvalues it is meaningful to include in 
our study. The crucial observation here is that in the microscopic
scaling region near $\lambda\sim 0$, the Dirac 
operator spectrum will have an average level spacing of
order unity. Specifically, we shall be concerned with the rescaled,
microscopic, spectral density \cite{SV}
\beq
\rho_s(\zeta) ~\equiv~ \frac{1}{V}~
\rho\left(\frac{\zeta}{V\Sigma}\right) ~,
\label{eq:rhoS}
\eeq
where $\Sigma= \langle\bar{\psi}\psi\rangle = \pi\rho(0)$, and $\rho(\lambda)$
denotes the usual (macroscopic) spectral density. We choose a
convention where the chiral condensate is defined, with a summation 
over the number of colors $N_c$, by (for one flavor of mass $m$)
\beq
\langle\bar{\psi}\psi\rangle ~=~ \frac{1}{V}\frac{\partial}{\partial m}
{\cal Z}(m) ~.
\label{eq:pbp}
\eeq
Here ${\cal Z}(m)$ is the finite-volume partition function.
The macroscopic spectral
density is normalized so that $\rho(\lambda)d\lambda$ is the mean number
of eigenvalues per unit volume, as in ref. \cite{LS}.
The exact, universal, analytical prediction from
both random matrix theory \cite{V,ADMN} and the finite-size partition
functions \cite{D,OTV} reads
\beq
\rho_s(\zeta) ~=~ \pi\rho(0)\frac{\zeta}{2}\left[J_{N_f+\nu}(\zeta)^2
-J_{N_f+\nu+1}(z)J_{N_f+\nu-1}(\zeta)\right] ~,
\label{eq:rho_s}
\eeq
with $\zeta\equiv\lambda V\Sigma=\pi\rho(0)V\lambda$.
This analytical expression has the following qualitative features: It 
vanishes at $\zeta=0$ (consistent with the fact that at any
finite volume there is no spontaneous chiral symmetry breaking),
rises sharply, and then undergoes exponentially damped oscillations
around the constant value $\rho(0)$\footnote{The microscopic spectral
density can be viewed as a blow-up of the point $\rho(0)$ due to the
finite volume. We should thus have lim$_{\zeta\to\infty}\rho_s(\zeta) = 
\rho(0)$, a matching condition which indeed is satisfied.}. The oscillations
take place on a scale of order $\Delta\zeta \sim 1$, and, in fact,
correspond directly to fluctuations over distinct sharp peaks
of the first few individual eigenvalues. While the analytical prediction
is exact (can be achieved to any degree of accuracy) in the 
infinite-volume limit with $L\ll
1/m_{\pi}$, it is clear that at any finite volume the prediction will
fail eventually. In any case, since the distinct feature of the
prediction, the oscillations, die out exponentially, we are in fact 
only interested in averaging over the first very few eigenvalues. It 
would thus be considerable overkill to compute, for each gauge field 
configuration, more than a few lowest-lying eigenvalues. We use the
Ritz functional method \cite{Ritz} to compute the lowest 10 eigenvalues (and
eigenvectors) of $-\Dslash^2$, with $\Dslash$ the staggered Dirac
operator
\beq
\Dslash_{x,y} = \frac{1}{2} \sum_\mu \eta_\mu(x) \left( U_\mu(x)
 \delta_{x+\mu,y} - U^\dagger_\mu(y) \delta_{x,y+\mu} \right)
 = \Dslash_{e,o} + \Dslash_{o,e} ~.
\label{eq:dslash}
\eeq
Here $\eta_\mu(x) = (-1)^{\sum_{\nu < \mu} x_\nu}$ are the usual staggered
phase factors. Introducing $\epsilon(x) = (-1)^{\sum_\nu x_\nu}$, we have
indicated that $\Dslash$ connects even, $\epsilon(x) = 1$, with odd,
$\epsilon(x) = -1$, sites and vice versa.

\noi
For clarity, we recall some well known properties of the staggered
Dirac operator here, see {\it e.g.} ref.~\cite{Kalkreuter}.
The staggered Dirac operator is antihermitian and therefore $-\Dslash^2$
is hermitian and positive (semi-) definite. $\epsilon(x)$ plays the r\^ole
that $\gamma_5$ plays for the continuum Dirac operator, {\it e.g.}
$\{\Dslash,\epsilon\} = 0$, and it is easy
to see that the eigenvalues of $\Dslash$ come in pairs $\pm i \lambda$.
If $\psi(x)$ is the eigenvector with eigenvalue $i \lambda$ then
$\epsilon(x) \psi(x)$ is the eigenvector with eigenvalue $-i \lambda$.
It is well known that $-\Dslash^2$ only couples even with even and
odd with odd sites. Furthermore, if $\psi_e$ is a normalized eigenvector of
$-\Dslash^2$ with eigenvalue $\lambda^2$, non-zero only on even sites,
then $\psi_o = \frac{1}{\lambda} \Dslash_{o,e} \psi_e$ is also  a
normalized eigenvector of $-\Dslash^2$ with eigenvalue $\lambda^2$,
non-zero only on odd sites.\footnote{Due to lattice artefacts the
staggered Dirac operator does not have exact zero-modes on equilibrium
gauge configurations at non-zero gauge coupling, $g^2$.}
The eigenvectors of $\Dslash$ with eigenvalues $\pm i \lambda$ are then
$\psi_{\pm}(x) = \frac{1}{\sqrt{2}} \left( \psi_e(x) \mp i \psi_o(x) \right)$.

\noi
In any case, it is clear that to obtain the lowest 10 positive (imaginary)
eigenvalues of $\Dslash$ it is sufficient to compute the lowest 10
eigenvalues of $-\Dslash^2_{e,e}$, working only on the even sublattice,
and then taking the square root. This is the approach we have taken here, 
for which the Ritz functional method is very well suited.

\noi
We have generated configurations that are essentially statistically independent
by employing a pure gauge algorithm of typically 3 to 4 microcanonical
overrelaxation sweeps, followed by one heatbath update sweep, and repeating
this process 10 times between eigenvalue computations. To convert our
eigenvalue measurements, or actually a histogram of all measured eigenvalues,
into a microscopic spectral density, see eq.~(\ref{eq:rhoS}), we need an
estimate of $\Sigma$, or of $\rho(0)$, in the infinite volume limit. On
a finite lattice $\rho(\lambda)$, obtained by normalizing a histogram
of the lowest eigenvalues appropriately, will oscillated about the infinite
volume $\rho(0)$. We thus obtain $\rho(0)$ just by averaging the finite
volume $\rho(\lambda)$ over the first few (leaving out the very first)
oscillations. Our results, together with information about the number of
measurements and the lattice size, is given in Table~\ref{tab:rho_0}.

\begin{table}
  \begin{center}
    \tabcolsep 4pt
    \vspace{2mm}
    \begin{tabular}{|c|c|c|c|c|} \hline
     $\beta$ & $L$ & meas & $\Delta \lambda$ & $\rho(0)$ \\ \hline
     4.0     &  4  & 5000 & 0.0007~ & 0.483(3)  \\
     4.5     &  4  & 4000 & 0.0008~ & 0.448(4)  \\
     5.1     &  4  & 4000 & 0.0010~ & 0.368(3)  \\
     5.1     &  6  & 4000 & 0.00018 & 0.368(3)  \\
     5.1     &  8  & 1500 & 0.00006 & 0.363(5)  \\
     5.55    &  6  & 4000 & 0.0003~ & 0.192(2)  \\
    \hline
    \end{tabular}
    \caption{Our estimates of $\rho(0)$ for our various gauge field
             ensembles. The third column gives the number of configurations
             analyzed, and the forth column the bin size in the histograms.}
    \label{tab:rho_0}
  \end{center}
  \vskip -5mm
\end{table}

\noi
Our results here are restricted to the quenched approximation, which, in
the analytical context can be achieved simply by taking the $N_f\to 0$ limit
of the formula (\ref{eq:rho_s}). While the quenched approximation obviously
saves an enormous amount of computer time, it also has another unique
advantage: The massless limit of the Dirac operator is taken automatically.
This means that the ``valence Goldstone boson'' in principle, were there
no lattice artefacts, would be strictly massless. One boundary of
the scaling regime, $L\ll 1/m_{\pi}$ is thus trivially satisfied from the very 
beginning. On the lattice, the other boundary, 
$1/\Lambda_{\mbox{\rm{\footnotesize QCD}}}
\ll L$, is effectively replaced by the condition that the volume should
be large in physical units. This means that the lattice volume in
lattice units must increase steeply with the gauge coupling $\beta=
6/g^2$ if we wish to go towards weak coupling, and continuum scaling.
Another way of saying that is as follows: to obtain reasonable estimates 
of $\rho(0)$ -- {\it i.e.} for $\rho(\lambda)$ to be approximately constant 
(apart from the predicted oscillations), we need sufficiently large lattices 
in physical units. Empirically we find that $L \gtwid 2 N_{\tau,c}(\beta)$ 
is needed, with $N_{\tau,c}(\beta)$ denoting the approximate temporal 
extension of a system to be at the deconfinement transition point at 
given gauge coupling $\beta$. For example, we found a $6^4$ lattice at 
$\beta=5.7$ ($N_{\tau,c}(5.7) \approx 4$) and an $8^4$ lattice at $\beta=5.85$
($N_{\tau,c}(5.85) \approx 6$) not to be sufficiently large.

\begin{figure}
\epsfxsize=7.0in
\centerline{\epsfbox[20 20 616 724]{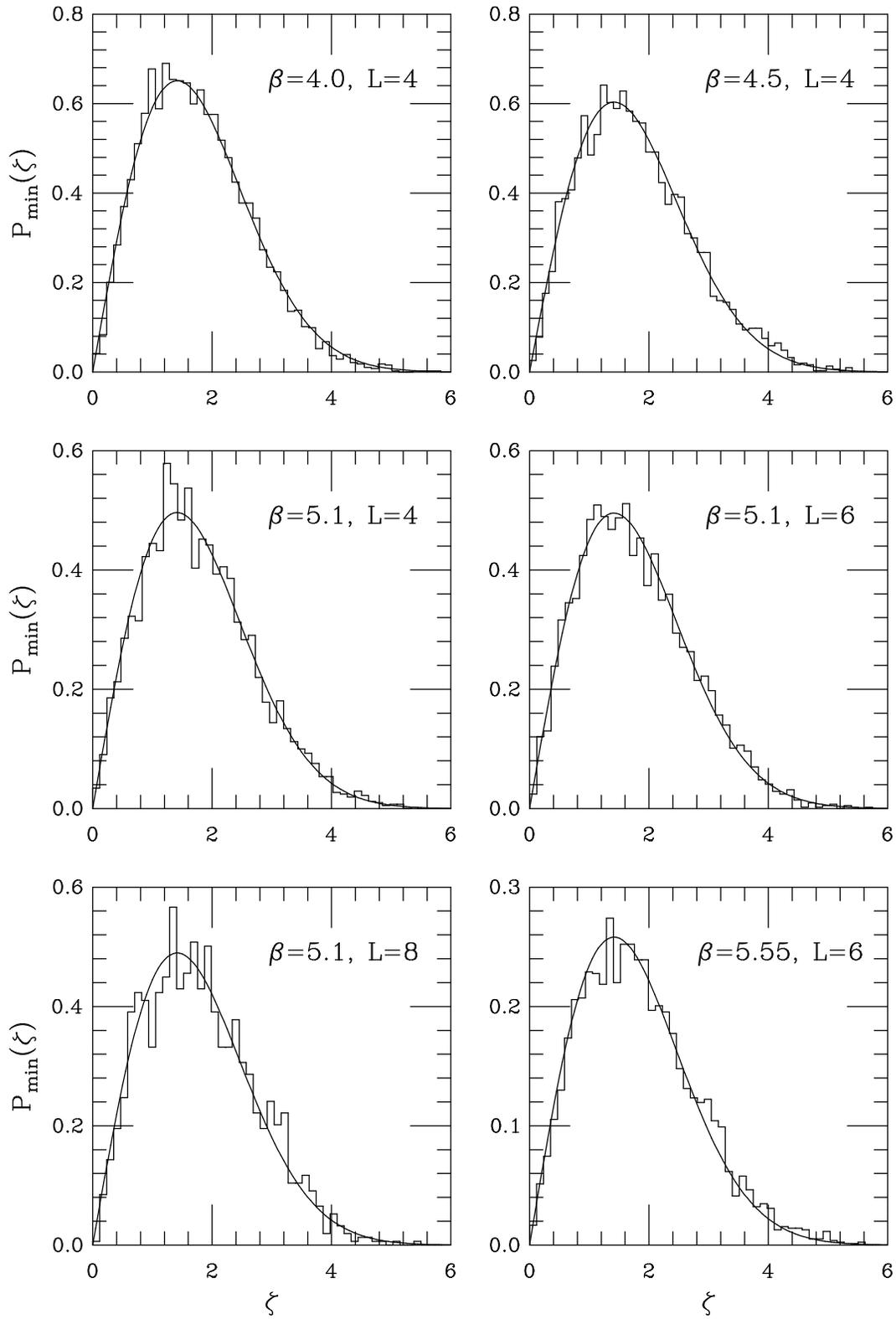}}
\caption{The distribution of the lowest eigenvalue for our different
         lattice ensembles of Table~\protect\ref{tab:rho_0}.}
\label{fig:Pmin}
\end{figure}

\noi
We begin our comparison between theory and Monte Carlo results with the
distribution of the smallest eigenvalue. Here the analytical prediction,
which again has been proven to be {\em universal} in the random matrix 
theory context, and derivable directly from the finite-volume partition 
functions too \cite{NDW}, reads
\beq
P_{\mbox{\rm{\footnotesize min}}}(\zeta) ~=~ \pi\rho(0)\frac{\zeta}{2}
\exp\left[-\zeta^2/4\right] ~.\label{eq:Pmin}
\eeq
Results for different lattice sizes of volume $L^4$ are shown in 
the histograms of Fig.~\ref{fig:Pmin}, which are compared with the
analytical prediction (\ref{eq:Pmin}). For each $\beta$-value we have
first determined $\rho(0)$ as described earlier, which means that these
plots are all parameter-free. The agreement is seen to be excellent indeed.

\begin{figure}
\epsfxsize=7.0in
\centerline{\epsfbox[20 20 616 724]{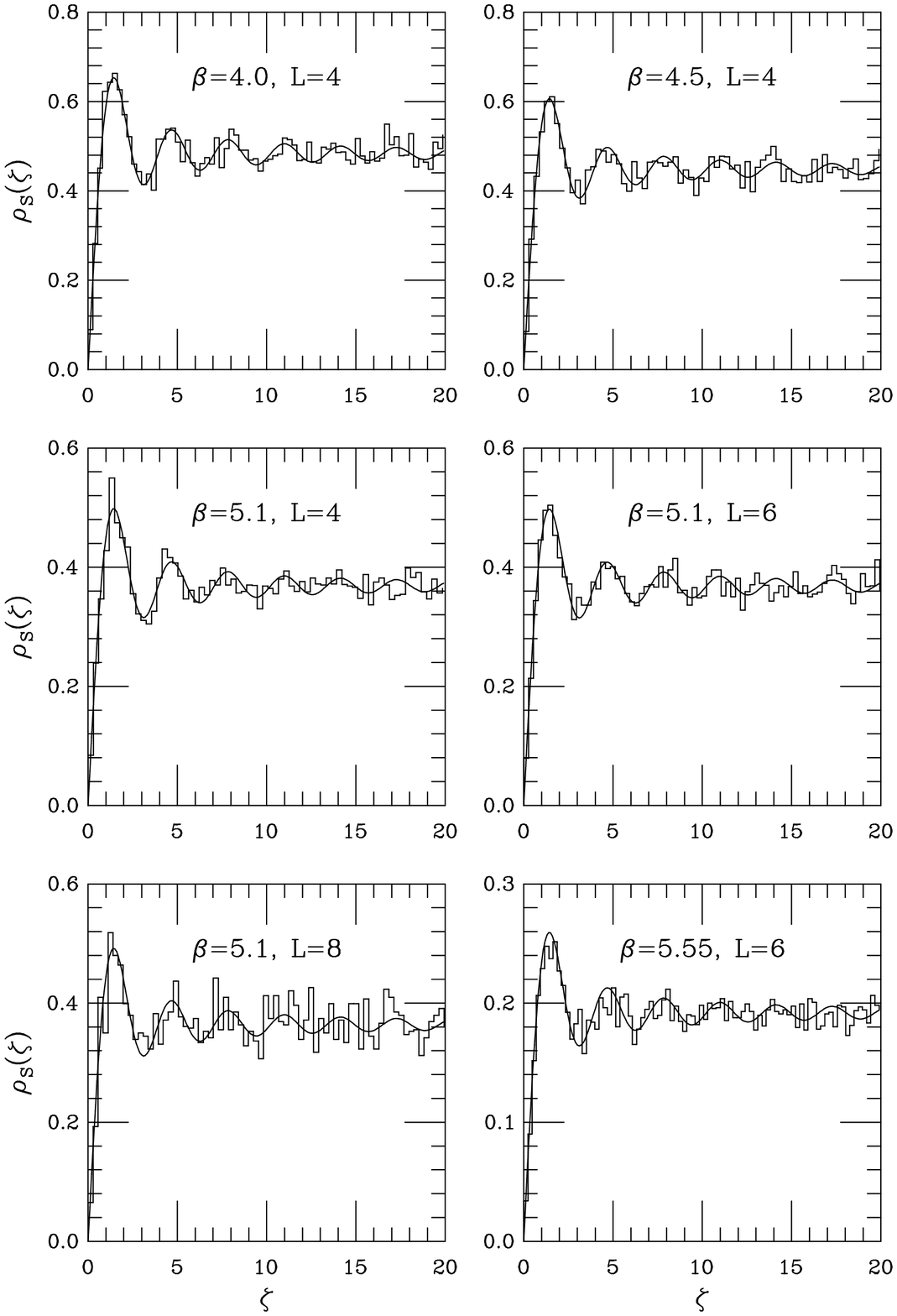}}
\caption{The microscopic spectral density for our different
         lattice ensembles of Table~\protect\ref{tab:rho_0}.}
\label{fig:rho_S}
\end{figure}

\noi
Next, we have extracted the spectral density $\rho(\lambda)$ near 
$\lambda \sim 0$ by binning eigenvalues in small intervals (see
Table~\ref{tab:rho_0}), and normalizing the resulting density as described
previously. By rescaling the small eigenvalues according to
$\zeta = \pi\rho(0)V\lambda$, we obtain the microscopic spectral density
as shown in Fig.~\ref{fig:rho_S}. Again we find absolutely excellent agreement
with the exact analytical prediction (\ref{eq:rho_s}), which is shown as
the full line. We have the poorest statistics for $\beta=5.1, L=8$ 
(only 1500 independent configurations), and indeed this is the plot
for which we get the least impressive agreement. However, taken together,
over several different $\beta$-values and several different volume
sizes, the precise prediction is very well reproduced. 

\noi
We conclude by noting that the exact analytical predictions for the
microscopic Dirac operator spectrum, and the smallest Dirac eigenvalue
distribution, now have been successfully confirmed by direct numerical
computation in SU(3) lattice gauge theory with staggered fermions. The
relevant universality class is, in the language of random matrix theory,
that of the chiral Unitary Ensemble.
For gauge group $SU(N_c)$ with $N_c \geq 3$ and fermions in the
fundamental representation, it coincides with the universality
class of the continuum Dirac operator.

\vspace{1cm}
\noindent
{\sc Acknowledgement:} P.H.D. and U.M.H. acknowledge the support of NATO
Science Collaborative Research Grant No. CRG 971487. The work of P.H.D has 
also been partially supported by EU TMR grant no. ERBFMRXCT97-0122, and the 
work of U.M.H. has been supported in part by DOE contracts DE-FG05-85ER250000
and DE-FG05-96ER40979. A.K. acknowledges 
the funding by the Portuguese Funda\c c\~ao para a Ci\^encia e a Tecnologia, 
grants CERN/S/FAE/1111/96 and CERN/P/FAE/1177/97.


\end{document}